\documentclass[preprint,amsmath,amssymb,floatfix]{revtex4-1}
\usepackage{graphicx}
\usepackage{dcolumn}
\usepackage{bm}
\usepackage{upgreek}
\usepackage{hyperref}
\usepackage[mathlines]{lineno}

\begin{document}
\title{Polygonal vortex beams in quasi-frequency-degenerate states}

\author{Yijie Shen}
\author{Zhensong Wan}
\author{Yuan Meng}
\author{Xing Fu} 
\author{Mali Gong} 
\email{shenyj15@mails.tsinghua.edu.cn}
\affiliation{State Key Laboratory of Precision Measurement Technology and Instruments, Department of Precision Instrument, Tsinghua University, Beijing 100084, China}

\date{\today}

\begin{abstract}
We originally demonstrate the vortex beams with patterns of closed polygons [namely polygonal vortex beams (PVBs)] generated by a quasi-frequency-degenerate (QFD) Yb:CALGO laser resonator with astigmatic transformation. The PVBs with peculiar patterns of triangular, square, and parallelogram shapes carrying large orbital angular momentums (OAMs) are theoretically investigated and experimentally obtained in the vicinity of the SU(2) degenerate states of laser resonator. The PVBs in QFD states are compared with the vortex beams with patterns of isolated spots arrays located on the triangle-, square-, and parallelogram-shaped routes [namely polygonal-spots-array vortex beams (PSA-VBs)] under normal SU(2) degenerate states. Beam profile shape of PVB or PSA-VB and OAM can be controlled by adjusting the cavity length and the position of pump spot. The simulated and experimental results validate the performance of our method to generate PVB, which is of great potential for promoting novel technologies in particle trapping and beam shaping.


\end{abstract}

\maketitle

\section{Introduction}
Optical vortex beams (OVBs) carrying orbital angular momentums (OAMs) are motivating widespread applications such as optical tweezers \cite{1,2,3}, optical communications \cite{4,5,6}, nanostructures processing \cite{7,8}, and quantum entanglement \cite{9,10}, to name a few. An OVB generally has a whirlpool-like phase distribution with the phase singularity giving rise to a dark hole at the center region of intensity pattern. The typical representatives are the classical doughnut-shaped Laguerre-Gaussian (LG) beams \cite{11,12}. Most recently, some peculiar OVBs with multiple vortices array \cite{13,14,15,16,17,18}, large fractional OAM \cite{19,20}, and specially tailored profiles \cite{21,22,23,24,25,26,27,28,29} other than the conventional doughnut-shaped LG beams are arousing great interest for exploiting novel theories and technologies. For instance, the Ince–Gaussian (IG) beams and helical-Ince-Gaussian (HIG) OVBs with hyperbolic or elliptical profile can be used in multi-particle manipulation \cite{1,13,14,15}. The Hermite-Laguerre-Gaussian (HLG) modes with an array of multiple singularities can be used to express mutual intensity evolution in closed form \cite{16,17,18}. G. Gbur recently proposed an intriguing fractional OVB with singularities of transfinite numbers, physically realizing the peculiar mathematical description of Hilbert’s Hotel paradox \cite{19,20}. L. Li et al. recently proposed a method to generate arrays of vortices along arbitrary curvilinear arrangement such as circle, square, pentagram, etc. which can be used in freely trapping of multiple particles \cite{21}. Y. F. Chen et al. has realized plenty of geometric modes with peculiar profiles of multiple circles \cite{22}, Lissajous parametric surfaces \cite{23,24}, quantum Green’s functions \cite{25,26}, and multiple spots shape \cite{27,28,29}, which all reveal a deeper quantum-classical connection as quantum coherent states in SU(2) Lie algebra \cite{30,31,32,33}. To sum up, the investigations upon novel structured vortex light fields are of great significance for frontier scientific researches.

In this work, we propose a new kind of polygonal vortex beams (PVB), which is generated by astigmatic transformation of the geometric modes in the vicinity of SU(2) coherent frequency-degenerate states. The PVBs carrying OAMs with patterns of closed triangle, square, and parallelogram shapes are obtained respectively in quasi-frequency-degenerate (QFD) states. The PVBs are distinct from the vortex beams with patterns of spots arrays located on the triangle-, square-, and parallelogram-shaped routes [namely polygonal-spots-array vortex beams (PSA-VBs)] in normal degenerate states as reported before \cite{27,28,29}. In order to obtain a stable QFD state, Yb:CALGO crystal is used as the gain medium in our resonator for its extremely broad emission band \cite{34,35,36}. The profile of PVB or PSA-VB as well as the OAM can be controlled by adjusting the cavity length and the position of pump spot. The new kind of PVB carrying OAM is of great potential for promoting novel technologies such as particle trapping and beam shaping.

\section{Theories}
\subsection{Frequency-degenerate resonator}
\begin{figure}
	\centering
	\includegraphics[width=0.7\linewidth]{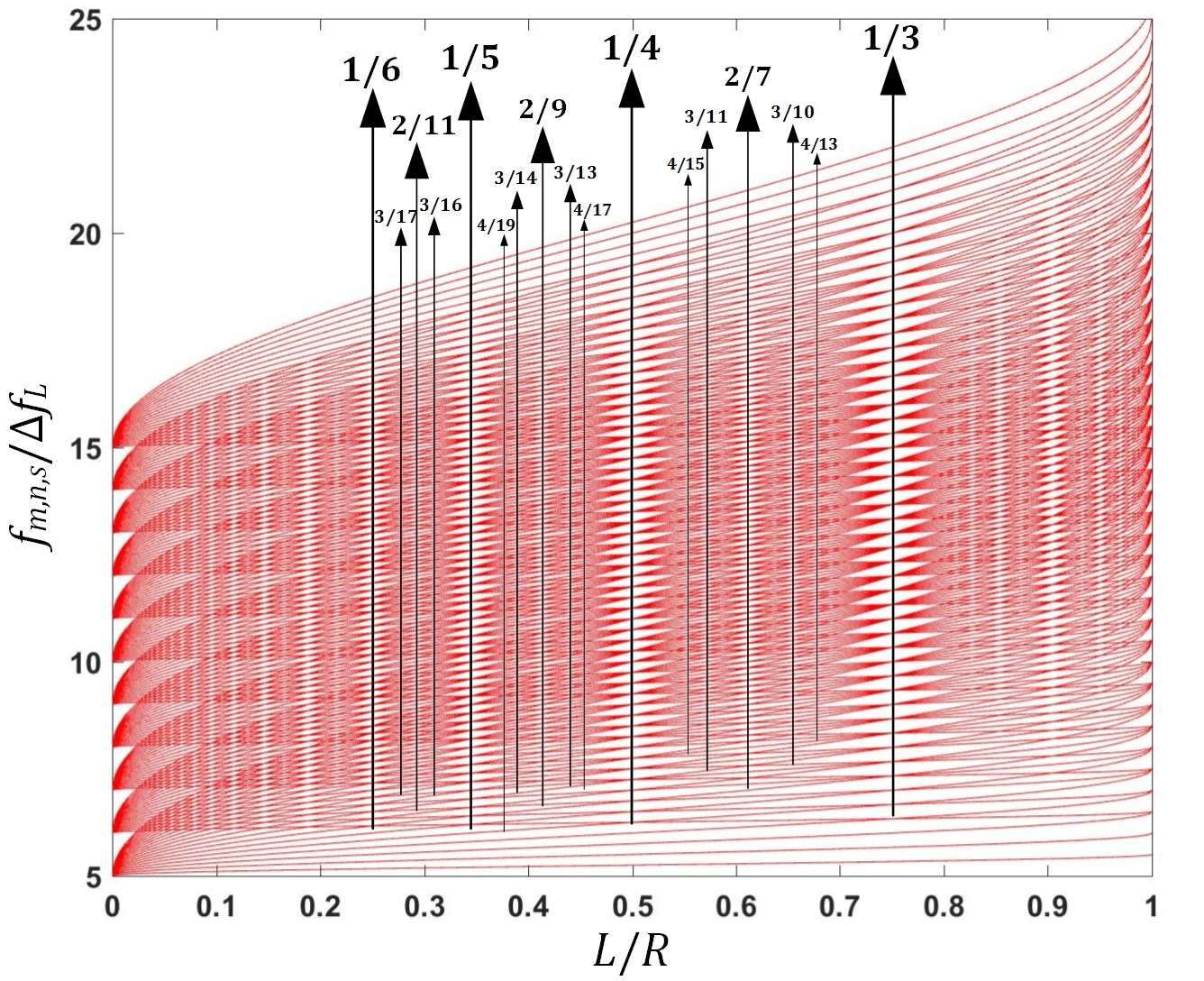}
	\caption{\label{f1} Diagram of the frequency-degenerate spectrum: $f_{n,m,s}/\Delta f_L$ as a function of the ratio $L/R$ for the range of $5\le s\le15$ and $0\le(m + n)\le20$.}
\end{figure}

To fulfill the reentrant condition of two-dimensional coupled harmonic oscillators in SU(2) Lie algebra, the cavity configuration requires a frequency-degenerate state of $\Omega=\Delta f_T/\Delta f_L=P/Q\in\mathbb{Q}$, where $P$ and $Q$ are coprime integers, and $\Delta f_L$ ($\Delta f_T$) is the longitudinal (transverse) mode spacing \cite{30,31,32,33}. The optical resonator satisfying a degenerate state is called frequency-degenerate cavity (FDC). Without loss of generality, we consider a plano-concave laser cavity with the length of $L$, comprised of a gain medium, a concave spherical mirror with the radius of curvature of $R$, and a planar output coupler \cite{30,31}. The eigenmodes $\psi_{n,m,s}$ and the eigenvalues $k_{n,m,s}$ for a laser cavity satisfy the Helmholtz equation:
\begin{equation}
\left( {{\nabla }^{2}}\text{+}k_{n,m,s}^{2} \right){{\psi }_{n,m,s}}\left( x,y,z \right)=0.
\end{equation}
Under the paraxial approximation, the eigenmodes that are separable in Cartesian coordinate $(x,y,z)$ for the cavity with a concave mirror at $z=-L$ and a plane mirror at $z=0$ can be expressed as the Hermite-Gaussian (HG) modes:
\begin{equation}
\psi _{n,m,s}^{\left( \text{HG} \right)}\left( x,y,z \right)=\sqrt{\frac{2}{L}}\cdot \Phi _{n,m}^{\left( \text{HG} \right)}\left( x,y,z \right)\exp \left[ \text{i}{{k}_{n,m,s}}\tilde{z}-\text{i}\left( m+n+1 \right){{\theta }_{G}}\left( z \right) \right],
\end{equation}
with the HG distribution:
\begin{equation}
\Phi _{n,m}^{\left( \text{HG} \right)}\left( x,y,z \right)=\frac{1}{\sqrt{{{2}^{m+n-1}}\pi m!n!}}\frac{1}{w\left( z \right)}{{H}_{n}}\left[ \frac{\sqrt{2}x}{w\left( z \right)} \right]{{H}_{m}}\left[ \frac{\sqrt{2}x}{w\left( z \right)} \right]\exp \left[ -\frac{{{x}^{2}}+{{y}^{2}}}{{{w}^{2}}\left( z \right)} \right],
\end{equation}
where $\theta_G(z)=\tan^{-1}(z/z_R)$ is the Gouy phase, $H_n(\cdot)$ represents the Hermite polynomials of $n$-th order, $k_{n,m,s}=2\pi f_{n,m,s}/c$, $f_{n,m,s}$ is the eigenmode frequency, $c$ is the speed of light, $\tilde{z}=z+(x^2+y^2)z/[2(z^2+z_R^2)]$, $w(z)=w_0\sqrt{1+(z/z_R )^2}$, $w_0=\sqrt{(\lambda z_R)/\pi}$ is the beam radius at the waist, and $\lambda$ is the emission wavelength. The eigenmode frequency of resonator can be given by $f_{n,m,s}=s\cdot\Delta f_L+(n+m+1)\cdot\Delta f_T$, where $s$ is the longitudinal mode index, $m$ and $n$ are the transverse mode indices. The longitudinal mode spacing is given by $\Delta f_L=c/(2L)$, where the minor disparity between the physical length and the geometric length is neglected. The transverse mode spacing is given by $\Delta f_T=\Delta f_L\theta_G(L)/\pi$. The mode-spacing ratio $\Omega=P/Q=(1/\pi)\cos^{-1}(\sqrt{1-L/R})$ reveals the degeneracy, which is varied in the range between 0 and 1/2 by changing the cavity length as $0<L<R$. The spectrum $f_{n,m,s}/\Delta f_L$ versus the ratio $L/R$ can illustrate the various degeneracy states distribution in FDC as topological joints of spectral lines. Fig.~\ref{f1} depicts a diagram of the frequency-degenerate spectrum in the range of $5\le s\le15$ and $0\le(m + n)\le20$, where some degeneracy states $|\Omega=P/Q\rangle$ are marked at corresponding positions. It has been recently proved that with simpler fraction $\Omega=P/Q$, the corresponding degenerate state is easier to be generated with higher intensity peak \cite{30}. Since the rational number field $\mathbb{Q}$ is dense and countable, the cavity parameters for the emergence of degenerate states can form a Devil’s Staircase \cite{25}. The irrational number field $(\mathbb{R}-\mathbb{Q})$ is uncountable, thus the non-degenerate states are generally more than the degenerate states based on mathematics of real analysis.

\subsection{Ray-wave duality}
When an optical resonator is operating close to FDC, its laser mode and intensity would undergo dramatic changes with the principle that laser modes have a preference to be localized on the periodic ray trajectories under off-axis pumping, which is called the ray-wave duality (RWD) \cite{30,31}.
Firstly, FDC mode has the ray-like property. Based on geometrical optics, the ABCD matrix is used to characterize the propagation property of the optical ray trajectories inside a stable plano-concave cavity \cite{37}. Since the cavity length satisfies $L/R=\sin^2(\Omega\pi)$ under degeneracy state $|\Omega=P/Q\rangle$, the corresponding ABCD matrix of the FDC is given by:
\begin{equation}
\mathbf{A}=\left( \begin{matrix}
1-\frac{2L}{R} & 2L\left( 1-\frac{L}{R} \right)  \\
-\frac{2}{R} & 1-\frac{2L}{R}  \\
\end{matrix} \right)=\left( \begin{matrix}
\cos \left( 2\Omega \pi  \right) & \frac{R}{2}{{\sin }^{2}}\left( 2\Omega \pi  \right)  \\
-\frac{2}{R} & \cos \left( 2\Omega \pi  \right)  \\
\end{matrix} \right).
\end{equation}
After $n$ times of round trips in the FDC, the matrix is derived as:
\begin{equation}
{{\mathbf{A}}^{n}}=\left( \begin{matrix}
\cos \left( 2n\Omega \pi  \right) & \frac{R}{2}{{\sin }^{2}}\left( 2n\Omega \pi  \right)  \\
-\frac{2}{R}\frac{\sin \left( 2n\Omega \pi  \right)}{\sin \left( 2\Omega \pi  \right)} & \cos \left( 2n\Omega \pi  \right)  \\
\end{matrix} \right).
\end{equation}
Since $Q$ is a prime number, $Q$-th power of \textbf{A} is a unit matrix:
\begin{equation}
{{\mathbf{A}}^{Q}}=\left( \begin{matrix}
\cos \left( 2Q\Omega \pi  \right) & \frac{R}{2}{{\sin }^{2}}\left( 2Q\Omega \pi  \right)  \\
-\frac{2}{R}\frac{\sin \left( 2Q\Omega \pi  \right)}{\sin \left( 2\Omega \pi  \right)} & \cos \left( 2Q\Omega \pi  \right)  \\
\end{matrix} \right)=\mathbf{I}.
\end{equation}
i.e., an optical ray oscillating at an arbitrary position within the cavity would coincide exactly with the initial state after $Q$ times of round trips. Therefore, it is proved that the lasing modes have a preference to be localized on the periodic ray trajectories in FDC.

Secondly, FDC mode has the wave-like property. Based on physical optics, the actual lasing mode is a certain paraxial structured Gaussian beam rather than simple optical rays. Deeply based on quantum optics, the lasing mode is characterized by coherent-state wave-function of a quantum harmonic oscillator. In common non-degenerate states, it has been confirmed that the high-order HG mode can be experimentally generated under off-axis pumping with a sufficiently large off-axis displacement \cite{35,38}. Whereas in degenerate states, it has been theoretically and experimentally confirmed that the emission mode can be characterized by the quantum coherent states with SU(2) Lie algebra \cite{33}:
\begin{equation}
\psi _{{{n}_{0}}}^{M}\left( x,y,z;{{\phi }_{0}}\left| \Omega  \right. \right)=\frac{1}{{{2}^{{M}/{2}\;}}}\sum\limits_{K=0}^{M}{\sqrt{\frac{M!}{K!\left( M-K \right)!}}}\cdot {{\text{e}}^{\text{i}K{{\phi }_{0}}}}\cdot \psi _{{{n}_{0}}+Q\cdot K,0,{{s}_{0}}-P\cdot K}^{\left( \text{HG} \right)}\left( x,y,z \right),
\label{e7}
\end{equation}
where $M+1$ stands for the number of HG modes in family of $\psi_{n_0+Q\cdot K,0,s_0-P\cdot K}^{(\text{HG})}(x,y,z)$ with $K=1,2,3,\cdots$, which constitute a family of frequency-degenerate modes, where $n_0$ and $s_0$ represent the minimum transverse and maximum longitudinal orders in the degenerate family respectively. Note that all the discussion corresponding to the off-axis pumping with $(\Delta x,0)$ can be applied to the case of $(0,\Delta y)$  for the degenerate family in terms of $\psi_{0,m_0+Q\cdot K,s_0-P\cdot K}^{(\text{HG})}(x,y,z)$. The wave-packet patterns of the coherent states $\psi_{n_0}^M(x,y,z;\phi_0)$ are verified to manifest the RWD, and the phase factor $\phi_0$ can be used to link to the starting ray of the geometric periodic orbits \cite{31}. When $M=0$ in Eq.~(\ref{e7}), the SU(2) coherent state is reduced to a HG mode, revealing the common non-degenerate states.

\begin{figure}
	\centering
	\includegraphics[width=0.8\linewidth]{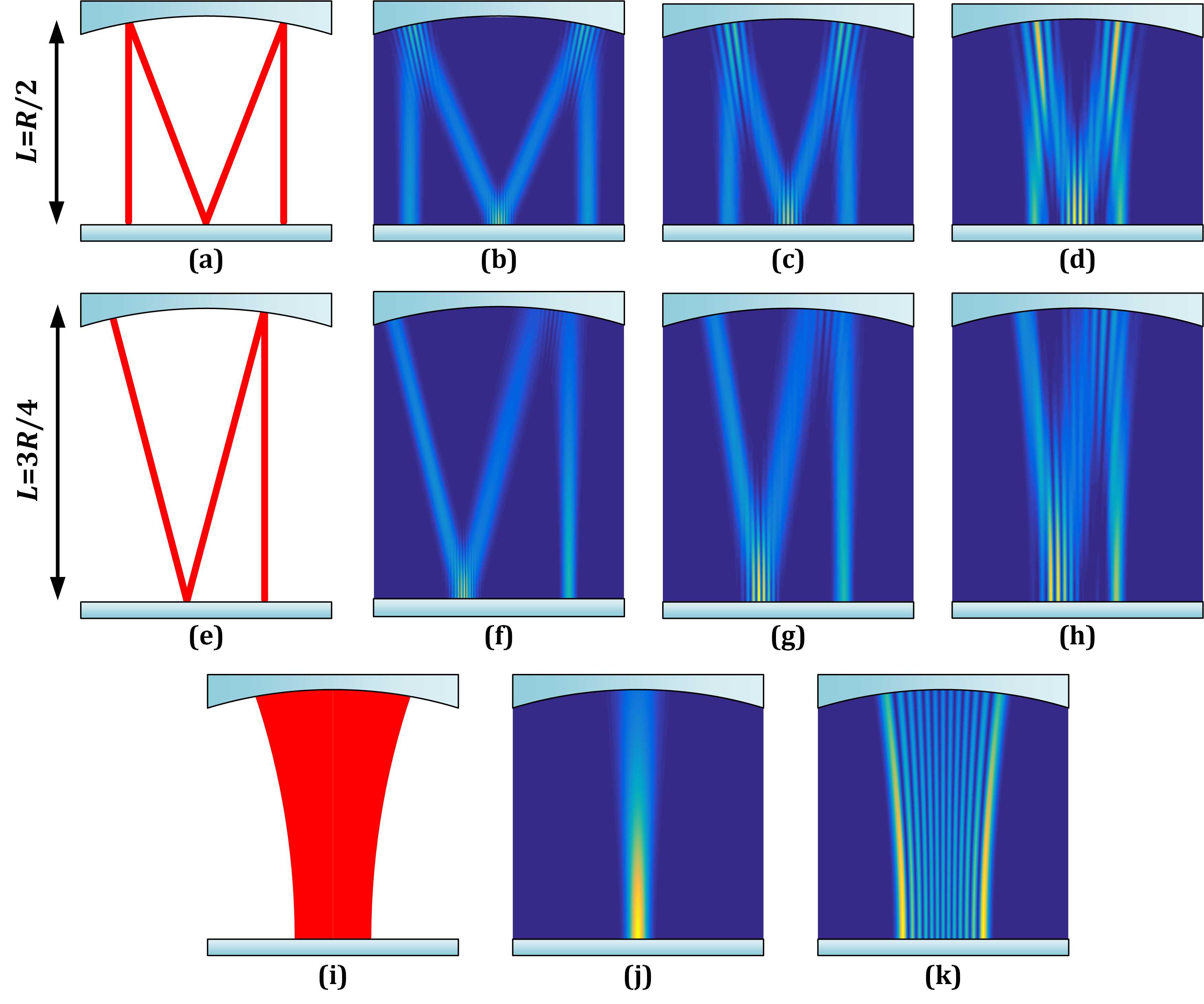}
	\caption{\label{f2} Theoretical periodic ray trajectories and SU(2) coherent-state modes in degenerate state (a-d) $|\Omega=1/4\rangle$, (e-h) $|\Omega=1/3\rangle$, and (i-k) non-degenerate state $|\Omega\in(\mathbb{R}-\mathbb{Q})\rangle$. (b-d) the intracavity patterns of $|\psi_{n_0}^M(x,0,z;0|\Omega=1/4)|^2$ with (b) $n_0=30$, $M=10$, (c) $n_0=12$, $M=5$, and (b) $n_0=7$, $M=2$. (f-h) the intracavity patterns of $|\psi_{n_0}^M (x,0,z;0|\Omega=1/3)|^2$ with (f) $n_0=30$, $M=10$, (g) $n_0=12$, $M=5$, and (h) $n_0=7$, $M=2$; (j) TEM$_{00}$ and (k) HG$_{12,0}$ modes in non-degenerate cavity.}
\end{figure}

Fig.~\ref{f2} depicts the theoretical periodic ray trajectories and SU(2) coherent-state modes under $|\Omega=1/4\rangle$ ($L=R/2$) [Fig.~\ref{f2}(a-d)], $|\Omega=1/3\rangle$ ($L=R/2$) [Fig.~\ref{f2}(e-h)], and $|\Omega\in(\mathbb{R}-\mathbb{Q})\rangle$ [Fig.~\ref{f2}(i-k)] with various combinations of $n_0$ and $M$. As can be seen in FDC, the mode with larger value of $n_0$ and $M$ is more analogous to show the ray property; the mode with smaller value of $n_0$ and $M$ is more analogous to exhibit the wave property. The degenerate state with a simpler rational fraction shows a more distinct RWD effect. In contrast, in non-degenerate cavity it is impossible to generate a periodic ray trajectories mode.

Therefore, FDC mode has the RWD.

\subsection{Quasi-frequency-degenerate states}
The frequency-degenerate theory is established under the assumption of monochromatic light field, while the actual laser beams are always quasi-monochromatic with a spectral linewidth. Thus, a mode in degenerate state cannot drastically collapse to a non-degenerate state when the FDC is adjusted to a condition in the vicinity of degenerate state. We propose the QFD wave-function to describe the mode behavior in the vicinity of FDC:
\begin{equation}
\widetilde{\psi }_{{{n}_{0}}}^{M}\left( x,y,z;{{\phi }_{0}}\left| \Omega  \right. \right)=\sum\limits_{K=0}^{M}{\left[ \frac{1}{{{2}^{{M}/{2}\;}}}\cdot \sqrt{\frac{M!}{K!\left( M-K \right)!}}\cdot {{\text{e}}^{\text{i}K{{\phi }_{0}}}}+{{q}_{K}} \right]}\cdot \psi _{{{n}_{0}}+Q\cdot K,0,{{s}_{0}}-P\cdot K}^{\left( \text{HG} \right)}\left( x,y,z \right),
\end{equation}
which can be seen as a superposition of a normal SU(2) degenerate modes and supplementary HG modes (HG mode means the non-degenerate state), where the sequence ${q_K}$ represent the weights of the superposed $\psi_{n_0+Q\cdot K,0,s_0-P\cdot K}^{\text{(HG)}}(x,y,z)$ modes. When ${q_K}$ satisfies $q_K=\delta_{Kj}-\text{e}^{\text{i}K{\phi_0}}\sqrt{M!/[2^MK!(M-K)!]}$, $j=0,1,\cdots,M$, where $\delta_{ij}$ is Kronecker symbol, the QFD mode is reduced to a HG mode $\psi_{n_0+Q\cdot K,0,s_0-P\cdot K}^{\text{(HG)}}(x,y,z)$.

\subsection{Astigmatic transformation of QFD modes}
With an external $\pi/2$-cylindrical-lens astigmatic mode converter (AMC), a HG mode with its principal axes inclined by 45$^\circ$ can be transformed into a doughnut-shaped LG mode \cite{11,12}. Using $\pi/2$-AMC, the QFD mode of quasi state $|\Omega=P/Q\rangle$ can be transformed into a vortex beam having pattern of the closed regular polygon with edges number of $Q$ (regular $Q$-gon). The transformed wave-function is written as:
\begin{eqnarray}
\widetilde{\Psi}{{_{{{n}_{0}}}^{M}}^{\left(\pm \right)}}\left( r,\varphi,z;{{\phi }_{0}}\left|\Omega \right. \right)=&&\sum\limits_{K=0}^{M}{\left[ \sqrt{\frac{M!}{K!\left( M-K \right)!}}\cdot \frac{{{\text{e}}^{\text{i}K{{\phi }_{0}}}}}{{{2}^{{M}/{2}\;}}}+{{q}_{K}} \right]}\cdot \Phi _{0,{{n}_{0}}+Q\cdot K}^{\left( \pm  \right)}\left( x,y,z \right) 
\nonumber\\&&\times \exp \left[ \text{i}{{k}_{{{n}_{0}}+Q\cdot K,0,{{s}_{0}}-P\cdot K}}\tilde{z}-\text{i}\left( m+n+1 \right){{\theta }_{G}}\left( z \right) \right], 
\end{eqnarray}
with the LG function:
\begin{equation}
\Phi _{p,\ell}^{\left(\pm \right)}\left(r,\varphi,z \right)=\sqrt{\frac{2p!}{\pi \left( p+\left| \ell\right| \right)!}}\frac{1}{w\left( z \right)}{{\left[ \frac{\sqrt{2}r}{w\left( z \right)} \right]}^{\left|\ell\right|}}L_{p}^{\left|\ell\right|}\left[ \frac{2{{r}^{2}}}{{{w}^{2}}\left( z \right)} \right]\exp \left[ -\frac{{{r}^{2}}}{{{w}^{2}}\left( z \right)} \right]\exp \left( \pm \text{i}\ell\varphi \right),
\end{equation}
where $(r,\varphi,z)$ represents the cylindrical coordinate, $L_{p}^{\left|\ell\right|}(\cdot)$ represents the associated Laguerre polynomials with radial and azimuthal indices of $p$ and $\ell$, and the sign of $\pm$ represents the chirality of vortex. For the case of $q_K=0$ (in degenerate state), the vortex beam has the pattern with the spots located on a $Q$-gon-shaped route. When the QFD cavity is adjusted to a non-degenerate state, the vortex beam tends to resemble a circular LG vortex beam.

For general AMC system, the Gouy phase difference between two orthogonal directions cannot always be strictly controlled as $\pi/2$ (i.e. the inclined angle $\alpha$ of input HG mode is not 45$^\circ$), the more general mode is in the form of the HLG mode with multiple singularities and fractional OAM \cite{16,17,18}. In general AMC systems, the corresponding transformed PVB with $Q$-gon profile for QFD mode can be written as:
\begin{eqnarray}
\widetilde{\Psi }{{_{{{n}_{0}}}^{M}}^{\left( \alpha  \right)}}\left( r,\varphi ,z;{{\phi }_{0}}\left| \Omega  \right. \right)=&&\sum\limits_{K=0}^{M}{\left[ \sqrt{\frac{M!}{K!\left( M-K \right)!}}\cdot \frac{{{\text{e}}^{\text{i}K{{\phi }_{0}}}}}{{{2}^{{M}/{2}\;}}}+{{q}_{K}} \right]}\cdot \Phi _{{{n}_{0}}+Q\cdot K,0,{{s}_{0}}-P\cdot K}^{\left( \text{HLG} \right)}\left( x,y,z\left| \alpha  \right. \right) \nonumber\\ 
&&\times\exp \left[ \text{i}{{k}_{{{n}_{0}}+Q\cdot K,0,{{s}_{0}}-P\cdot K}}\tilde{z}-\text{i}\left( m+n+1 \right){{\theta }_{G}}\left( z \right) \right],
\end{eqnarray}
with the HLG function:
\begin{eqnarray}
\Phi _{n,m,s}^{\left( \text{HLG} \right)}\left( x,y,z\left| \alpha  \right. \right)=&&\frac{1}{\sqrt{{{2}^{m+n-1}}m!n!}}\cdot \exp \left[ -\frac{{{x}^{2}}+{{y}^{2}}}{{{w}^{2}}\left( z \right)} \right] \nonumber\\ 
&& \text{ }\times {{\left. \left\{ \frac{{{\partial }^{m}}}{\partial s_{x}^{m}}\frac{{{\partial }^{n}}}{\partial s_{y}^{n}}\exp \left[ -{{\left( {{\mathbf{U}}^{*}}\mathbf{s} \right)}^{\text{T}}}\left( \mathbf{Us} \right)+\frac{2\sqrt{2}}{w\left( z \right)}{{\left( {{\mathbf{U}}^{*}}\mathbf{s} \right)}^{\text{T}}}\left( \begin{matrix}
		x  \\
		y  \\
		\end{matrix} \right) \right] \right\} \right|}_{\mathbf{s}=\mathbf{0}}},
\end{eqnarray}
where $\mathbf{s}={{\left( {{s}_{x}},{{s}_{y}} \right)}^{\text{T}}}$ and $\mathbf{U}=\left( \begin{matrix}
\cos \alpha  & \text{i}\sin \alpha   \\
\text{i}\sin \alpha  & \cos \alpha   \\
\end{matrix} \right)$. When $\alpha = 0$ or $\pi/2$, the HLG mode is reduced to HG mode and the transformed mode $\widetilde{\Psi }_{n_0}^{M(\alpha)}(x,y,z;\phi_0)$ is reduced to QFD mode $\widetilde{\psi}_{n_0}^{M}(x,y,z;\phi_0)$. When $\alpha=\pm\pi/4$, the HLG mode is reduced to LG mode, and $\widetilde{\Psi }_{n_0}^{M(\alpha)}(x,y,z;\phi_0)$ is reduced to PVB mode $\widetilde{\Psi}_{n_0}^{M(\pm)}(x,y,z;\phi_0)$ with regular $Q$-gon profile. More detailed morphologies (e.g. singularities distribution and intensity profile) of the vortex beam with the closed $Q$-gon pattern are dominated by ${q_K}$, which controls the interpolated process between the $Q$-gon profile and the circular profile.

\section{Experimental setup}
\begin{figure}
	\centering
	\includegraphics[width=0.95\linewidth]{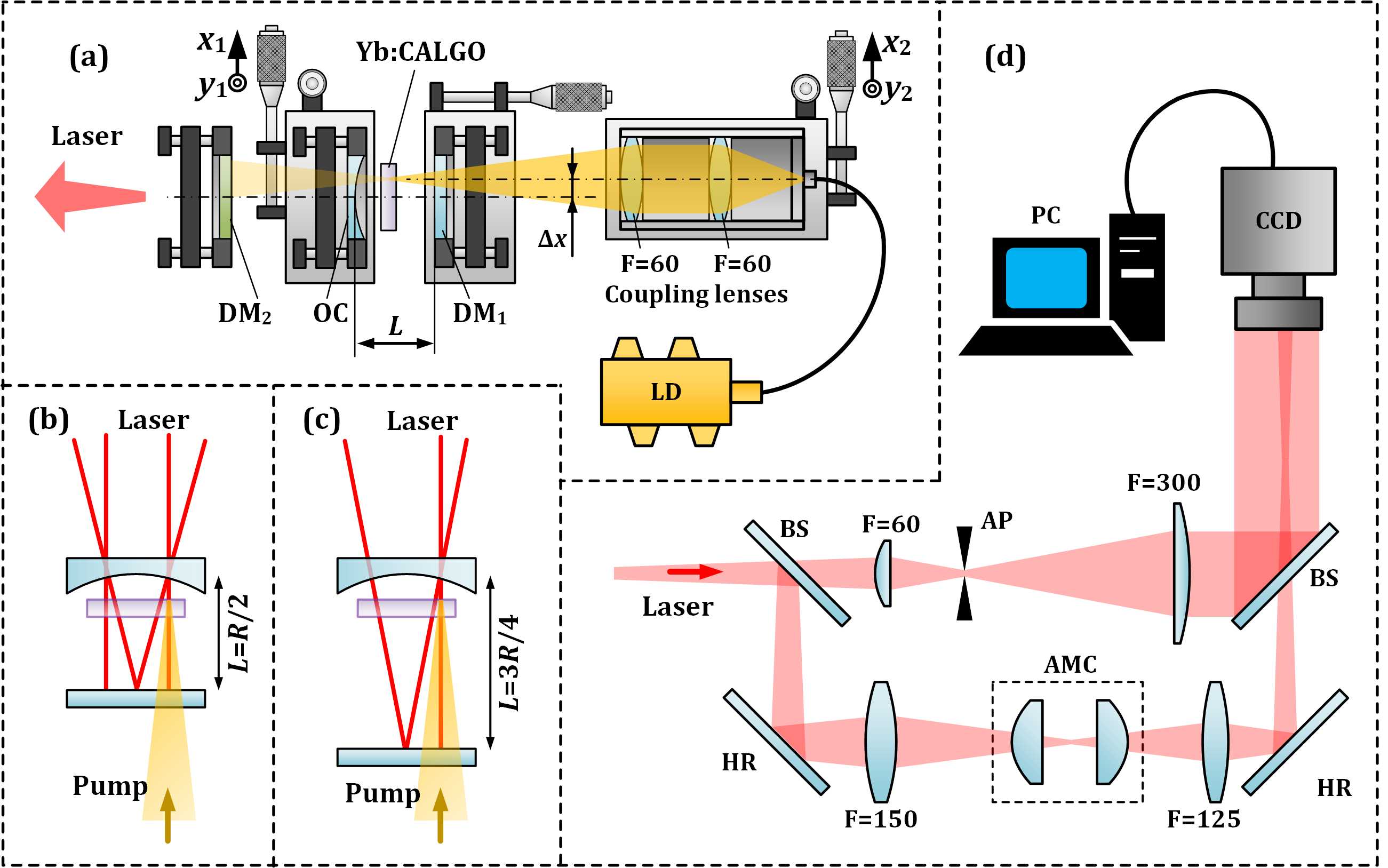}
	\caption{\label{f3} The experimental setup: (a) the diode-pumped laser oscillator; schematics of the degenerate cavities in degenerate states of (b) $|\Omega=1/4\rangle$ and (c) $|\Omega=1/3\rangle$; (d) the Mach-Zehnder interferometer for generating and measuring OAM. LD, laser diode; DM, dichroic mirror; OC, output coupler; AMC, astigmatic mode converter; HR, high-reflective mirror; BS, beam splitter; AP, aperture; CCD, charge coupled device; PC, personal computer.}
\end{figure}

The experimental setup of the laser oscillator is shown in Fig.~\ref{f3}(a). A high-power 976 nm fiber-coupled laser diode (LD) (Han’s TCS, core: 105 $\upmu$m, NA: 0.22, highest power: 110 W) was used as the pump source. A $4\times4\times×2$ mm$^3$ a-cut Yb:CALGO (Altechna, 5 at.\%) is used as the gain medium. Taking advantage of the extremely broad emission spectrum of Yb:CALGO, the laser crystal possesses ameliorated quasi-monochromatic property compared with common narrow-linewidth crystal (e.g. Nd:YAG and Nd:YVO$_4$). Therefore, a stable QFD state is easier to be realized. With two identical antireflective (AR) coated lenses with the focal length of $F=60$ mm, the pump light was focused into the crystal with the beam waist radius of about 200 $\upmu$m. The end surfaces of Yb:CALGO were AR coated at 900–1100 nm, which was wrapped in a copper heat sink water-cooled at 18$^\circ$C. The linear cavity included a plane mirror [DM$_1$: AR at 976 nm and high-reflective (HR) at 1030-1080 nm] and a concave output coupler (OC, with the transmittance of 1\% at 1020-1080 nm; $R=100$~mm). Another dichroic mirror (DM$_2$: HR at 976 nm and AR at 1030-1080 nm) was used to filter the residual pump light. All the coupling lenses system, OC, and DM$_1$ are installed on three-dimensional high-precision displacement stages, thus the off-axis displacement of pumping and the cavity length can be precisely adjusted. In our setup, the adjustable range of cavity length can cover the two degenerate states $|\Omega=1/4\rangle$ and $|\Omega=1/3\rangle$ at $L = R/2 = 50$ mm and $L = 3R/4 = 75$ mm, as shown in Figs.~\ref{f3}(b) and (c) respectively. 

Moreover, A Mach-Zehnder interferometer was built for generating and measuring the OAM, as shown in Fig.~\ref{f3}(d), the two arms were formed by two beam splitters and two 45$^\circ$ HR mirrors. For the one arm, the laser was incident into the AMC after being focused by a convex lens ($F = 150$ mm), and then be captured by a CCD camera (Spiricon, M2-200s) after being focused by another convex lens ($F = 125$ mm). The AMC constituted by two identical convex-plane cylindrical lenses with the focal length of $f = f_1 = f_2 = 25$ mm and a separation of 35.4 mm ($\sqrt{2}f$). For the another arm, a confocal telescope including two convex lenses ($F = 60$ mm and $F = 300$ mm) with an aperture in between was used to convert the beam into a near plane wave, which is used to obtain interference diagram for verifying the OAM.

\section{Results and discussions}
\subsection{Generation of QFD modes}
\begin{figure}
	\centering
	\includegraphics[width=\linewidth]{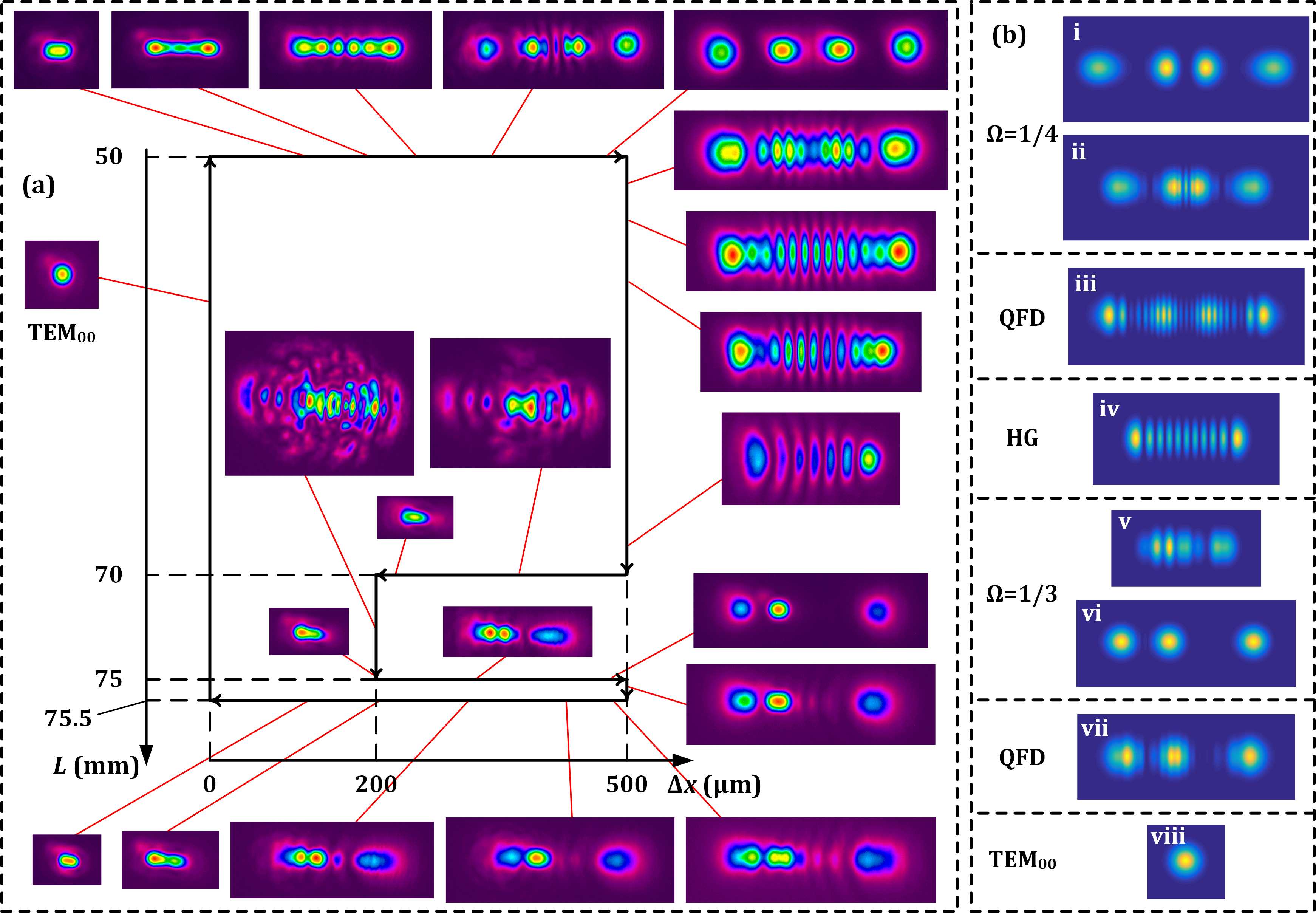}
	\caption{\label{f4} (a) The experimental emission modes evolution versus various $L$ and $\Delta x$, and (b) some simulated intensity patterns in selected states for references: 
		(i) $|\psi_5^{12} (x,y,0;0.3\pi|\Omega=1/4)|^2$, 
		(ii) $|\psi_4^6 (x,y,0;0.25\pi|\Omega=1/4)|^2$, 
		(iii) $|\widetilde{\psi}_2^8 (x,y,0;0.35\pi|\Omega=1/4)|^2$ when $q_K=0.8\cdot\delta_{K,4}$, (iv) HG$_{10,0}$, (v) $|\psi_2^2 (x,y,0;0.35\pi|\Omega=1/3)|^2$, 
		(vi) $|\psi_8^{12} (x,y,0;0.35\pi|\Omega=1/3)|^2$, 
		(vii) $|\widetilde{\psi}_3^8 (x,y,0;0.4\pi|\Omega=1/3)|^2$ 
		when $q_K=0.5\cdot\delta_{K,4}$, 
		(viii) TEM$_{00}$.}
\end{figure}
Fig.~\ref{f4}(a) shows the evolution of the experimentally measured emission modes versus various $L$ and $\Delta x$. With zero off-axis displacement, the output mode is near-TEM$_{00}$ for various $L$. For $L = 50$ mm, the cavity is in $|\Omega=1/4\rangle$, and RWD effect is getting more obvious as Δx gradually increased. At $\Delta x = 0.5$ mm, a pattern with four bright isolated spots is obtained involved in a periodic ray orbit in FDC. When $L$ is enlarged, the degenerate mode gradually changes into QFD state and then tends to a HG-like pattern. In this process, the laser intensity gradually decreases until the emission mode vanishes around $L = 70$ mm. With the off-axis displacement shortened to $\Delta x = 0.2$ mm, the intensity slightly recovers but many mixed transverse patterns emerge. Afterwards, when $L$ is increased to 75 mm (corresponding to the state $|\Omega=1/3\rangle$), the mixed mode becomes organized, and the periodic ray-orbit mode with period of three becomes more obvious with the increment of $\Delta x$, following the principle of RWD. We also slightly adjust the cavity length around $L = 75$ mm to observe the mode morphology of QFD states. For the modes evolution in $L = 75.5$ mm, a QFD state in the vicinity of $|\Omega=1/3\rangle$), some exotic strips are mixed in the QFD mode. It should be noted that we have not observed any periodic ray-orbit degenerate mode between $|\Omega=1/3\rangle$ and $|\Omega=1/4\rangle$, such as $|\Omega=2/7\rangle$, $|\Omega=3/11\rangle$, and $|\Omega=3/10\rangle$, in contrast to the previous results based on the narrow-linewidth Nd:YVO$_4$ where more fractional degenerate states can also be generated. Fig.~\ref{f4}(b) shows the simulated interpretations for the experimental degenerate and QFD modes according to the QFD theories in the previous section. 

\subsection{Generation of polygonal vortex beams}
\begin{figure}
	\centering
	\includegraphics[width=0.7\linewidth]{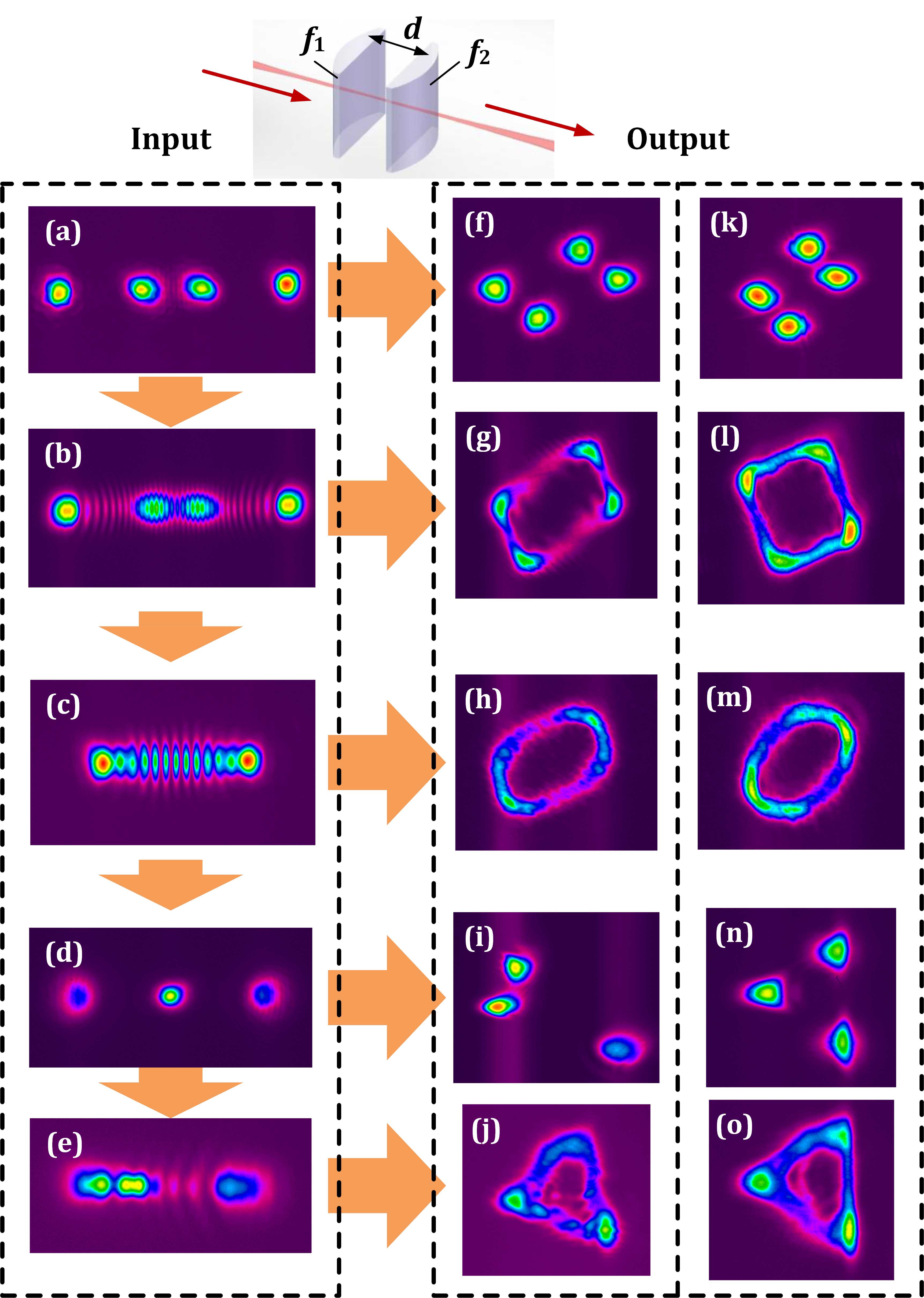}
	\caption{\label{f5} The experimental results of mode transformation of (a-e) various degenerate, quasi-degenerate non-degenerate modes, (f-j) the corresponding transformed modes by AMC, and (k-o) the corresponding transformed modes by near-$\pi/2$-AMC.}
\end{figure}

The beam transformation with the AMC is an effective method to generate optical vortex beams carrying OAM via converting HG modes into the general HLG modes \cite{17,18}, as well as transforming planar geometric modes into circular geometric modes \cite{28}. Hereinafter, we demonstrate that the QFD modes can be transformed into PVBs carrying OAM. Fig.~\ref{f5} shows the experimental results of mode transformation of the modes in (a) $|\Omega=1/4\rangle$, (b) quasi state $|\Omega\approx1/4\rangle$, (c) non-degenerate, (d) $|\Omega=1/3\rangle$, and (e) quasi state $|\Omega\approx1/3\rangle$ states. For $|\Omega=1/4\rangle$, the transformed PSA-VB mode shows a spots-array pattern along a parallelogram-shaped route after a general AMC. Especially, the spots array can be gradually located on a square-shaped route when the AMC is near-$\pi/2$-AMC condition (the inclined angle $\alpha$ of input HG mode is adjusted to 45$^\circ$). For the quasi state $|\Omega\approx1/4\rangle$, the transformed PVB mode shows a closed parallelogram shape for AMC and especially a square shape for near-$\pi/2$-AMC. For non-degenerate state, the transformed mode is elliptical for AMC and tends to a circle for near-$\pi/2$-AMC. For $|\Omega=1/3\rangle$, the transformed PSA-VB modes show an spots-array pattern along triangular route for AMC and especially the spots array can be located on a regular triangle route for near-$\pi/2$-AMC. For the quasi state $|\Omega\approx1/3\rangle$, the transformed modes show a closed triangular pattern for AMC and especially a closed regular-triangle-shaped pattern for near-$\pi/2$-AMC. 

\begin{figure}
	\centering
	\includegraphics[width=\linewidth]{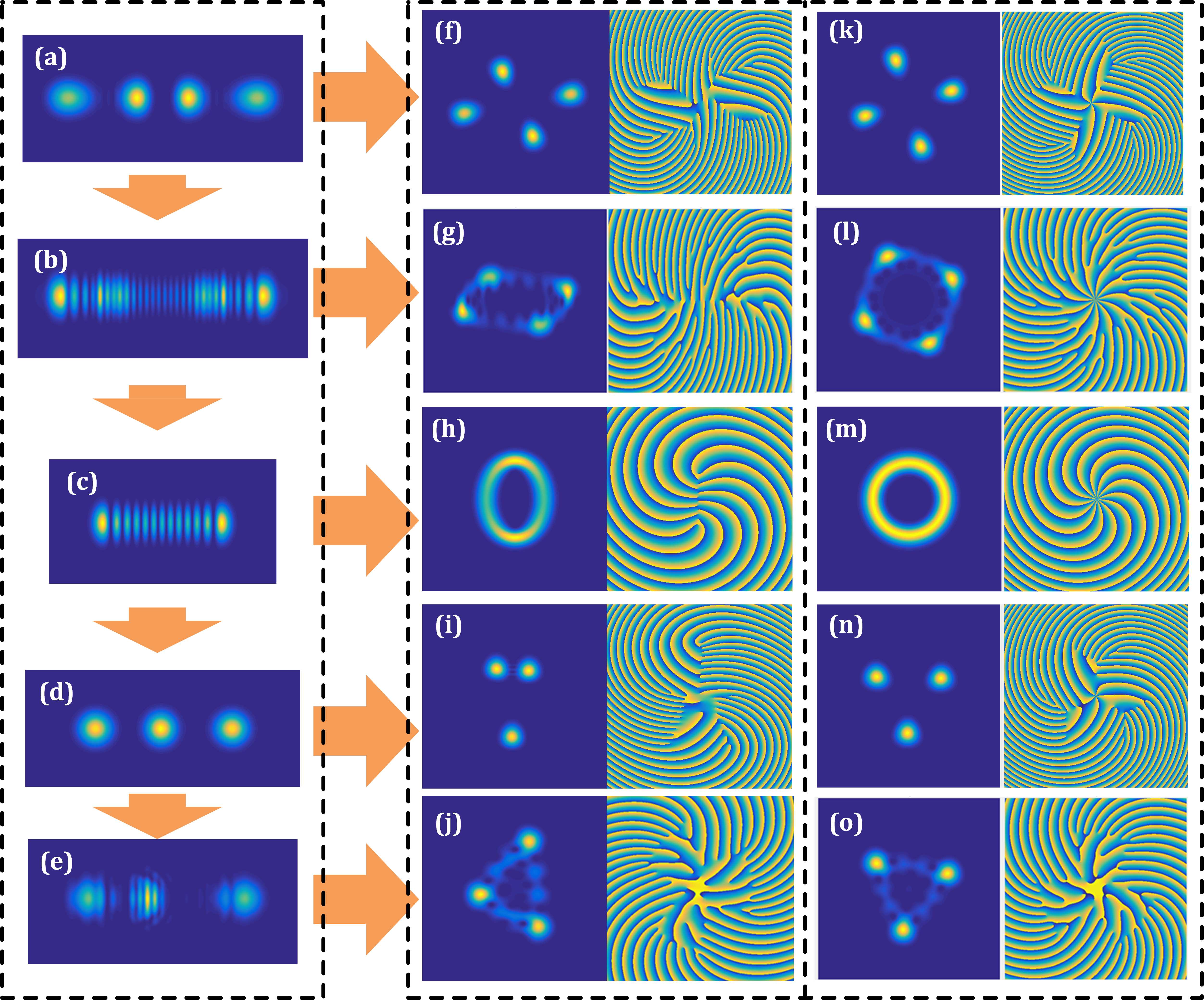}
	\caption{\label{f6} The simulated intensity patterns with vortex phase distribution as the interpretations of the experimental modes in Fig.~\ref{f5}.}
\end{figure}

Fig.~\ref{f6} shows the simulated results of the transformed modes, which are in good agreement with the experimental results. The minor aberrations can be further comprehended by the experimental misalignment of optical path in Mach-Zehnder interferometer setup. The simulated interpretations for the corresponding above-mentioned transformed modes are given by (a) $|\psi_5^{12}(x,y,0.8z_R;\pi/2|\Omega=1/4)|^2$, (b) $|\widetilde{\psi}_{10}^9(x,y,0.8z_R;0|\Omega=1/4)|^2$, when $q_K=0.4(u_K^{(9)}\text{e}^{-\text{i}K\pi/2}+u_K^{(6)}\text{e}^{-\text{i}K\pi}+u_K^{(4)}\text{e}^{-\text{i}K\cdot3\pi/2})$, where $u_K^{(N)}=1$ for $K\le N$, or 0 for $K > N$. (c) $|\psi_{12}^0(x,y,0.8z_R)|^2$, (d) $|\psi_5^{10}(x,y,0.8z_R;\pi/3|\Omega=1/3)|^2$, (e) $|\widetilde{\psi}_3^{13}(x,y,0.8z_R;0|\Omega=1/3)|^2$, when $q_K=0.4u_K^{(6)}(\text{e}^{-\text{i}K\cdot2\pi/3}+\text{e}^{-\text{i}K\cdot4\pi/2})$, (f) $|{\Psi}_5^{12(0.175\pi)}(x,y,0.8z_R;\pi/2|\Omega=1/4)|^2$, (g) $|\widetilde{\Psi}_{10}^{9(0.14\pi)}(x,y,0.8z_R;0|\Omega=1/4)|^2$, (h) $|{\Psi}_{12}^{0(0.3\pi)}(x,y,0.8z_R)|^2$, (i) $|{\Psi}_5^{10(3\pi/8)}(x,y,0.8z_R;\pi/3|\Omega=1/3)|^2$, (j) $|\widetilde{\Psi}_3^{13(0.28\pi)}(x,y,0.8z_R;0|\Omega=1/3)|^2$, (k) $|{\Psi}_5^{12(+)}(x,y,0.8z_R;\pi/2|\Omega=1/4)|^2$, (l) $|\widetilde{\Psi}_{10}^{9(+)}(x,y,0.8z_R;0|\Omega=1/4)|^2$, (m) $|{\Psi}_{12}^{0(+)}(x,y,0.8z_R)|^2$, (n) $|{\Psi}_5^{10(+)}(x,y,0.8z_R;\pi/3|\Omega=1/3)|^2$, (o) $|\widetilde{\Psi}_3^{13(+)}(x,y,0.8z_R;0|\Omega=1/3)|^2$. The basic principle of mode evolution of PVBs is revealed by both simulated and experimental results.

\begin{figure}
	\centering
	\includegraphics[width=\linewidth]{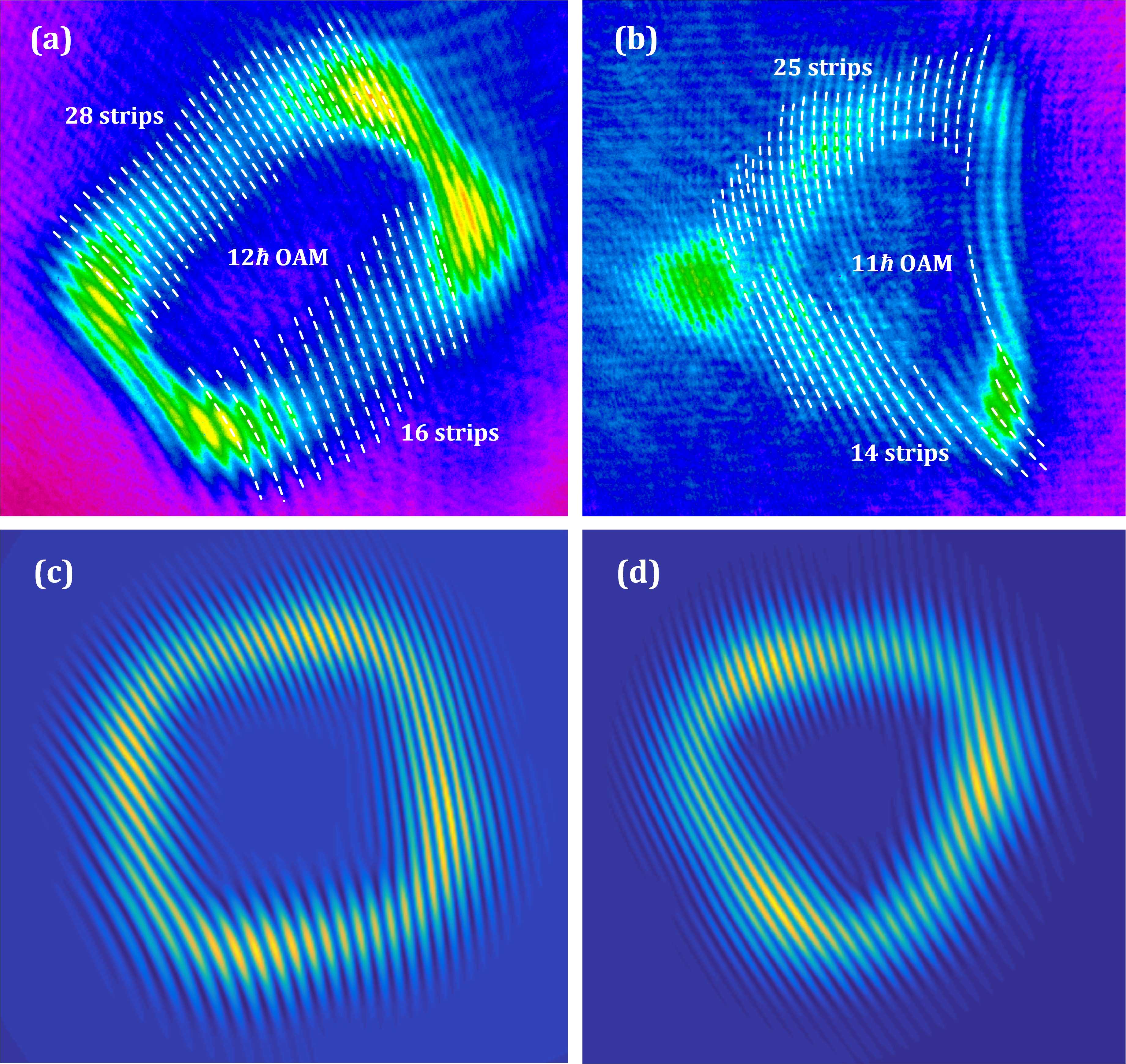}
	\caption{\label{f7} The (a,b) experimental and (c,d) simulated results of interference patterns of (a,c) square and (b,d) triangular vortex beams for verifying the OAM.}
\end{figure}

Figs.~\ref{f7}(a) and (b) show the experimental interference patterns of the square and triangular vortex beams with the coherent near-plane-wave, showing OAMs in the dark regions of $12\hbar$ and $11\hbar$ per photon respectively. Figs.~\ref{f7}(c) and (d) show the corresponding theoretical interpretations by using QFD modes $|\widetilde{\Psi}_5^4 (x,y,0.8z_R;0.6\pi|\Omega=1/4)|^2$ when $q_K=(1-|K/M-0.5|)^6\cdot\delta_{K,4}$ and $|\widetilde{\Psi}_4^4 (x,y,0.8z_R;0.6\pi|\Omega=1/3)|^2$ when $q_K=(1-|K/M-0.5|)^5\cdot\delta_{K,4}$, respectively.

\section{Conclusion}
In summary, we propose a method to generate polygonal optical vortices carrying OAM with pattern in the closed $Q$-gon shapes by transforming QFD modes in the vicinity of degenerate state $|\Omega=P/Q\rangle$ via AMC. A Yb:CALGO crystal is used as the gain medium in our resonator to obtain a stable quasi-degenerate state, taking advantage of its extremely broad emission band. The PVB with closed triangle shape is obtained at QFD state in the vicinity of ├ $|\Omega=1/3\rangle$, in contrast to the PSA-VB with spots-array pattern along triangular route in the normal degenerate state. The PVB with closed square and parallelogram shape are respectively obtained at QFD state in the vicinity of $|\Omega=1/4\rangle$, in contrast to the PSA-VB with spots-array patterns along corresponding square- and parallelogram-shaped route in the normal degenerate state. Besides, more $Q$-gon vortex beams with $Q\ge5$ are not exhibited due to the limitation of actual adjustable range of our experimental setup, and the realization and physical origin of which will be one of the focuses in further works. The simulated and experimental results show good agreement for validating the performance of the novel vortex beams with closed polygon pattern, which is of great potential to inspire novel technologies in particle trapping and beam shaping.

\end{document}